\documentstyle[11pt,psfig,twoside]{article}

\begin{document}

\newcommand{\dd}{\,{\rm d}}
\newcommand{\ie}{{\it i.e.},\,}
\newcommand{\etal}{{\it et al.\ }}
\newcommand{\eg}{{\it e.g.},\,}
\newcommand{\cf}{{\it cf.\ }}
\newcommand{\vs}{{\it vs.\ }}
\newcommand{\zdot}{\makebox[0pt][l]{.}}
\newcommand{\up}[1]{\ifmmode^{\rm #1}\else$^{\rm #1}$\fi}
\newcommand{\dn}[1]{\ifmmode_{\rm #1}\else$_{\rm #1}$\fi}
\newcommand{\upd}{\up{d}}
\newcommand{\uph}{\up{h}}
\newcommand{\upm}{\up{m}}
\newcommand{\ups}{\up{s}}
\newcommand{\arcd}{\ifmmode^{\circ}\else$^{\circ}$\fi}
\newcommand{\arcm}{\ifmmode{'}\else$'$\fi}
\newcommand{\arcs}{\ifmmode{''}\else$''$\fi}
\newcommand{\MS}{{\rm M}\ifmmode_{\odot}\else$_{\odot}$\fi}
\newcommand{\RS}{{\rm R}\ifmmode_{\odot}\else$_{\odot}$\fi}
\newcommand{\LS}{{\rm L}\ifmmode_{\odot}\else$_{\odot}$\fi}

\newcommand{\Abstract}[2]{{\footnotesize\begin{center}ABSTRACT\end{center}
\vspace{1mm}\par#1\par
\noindent
{~}{\it #2}}}

\newcommand{\TabCap}[2]{\begin{center}\parbox[t]{#1}{\begin{center}
  \small {\spaceskip 2pt plus 1pt minus 1pt T a b l e}
  \refstepcounter{table}\thetable \\[2mm]
  \footnotesize #2 \end{center}}\end{center}}

\newcommand{\TableSep}[2]{\begin{table}[p]\vspace{#1}
\TabCap{#2}\end{table}}

\newcommand{\FigCap}[1]{\footnotesize\par\noindent Fig.\  %
  \refstepcounter{figure}\thefigure. #1\par}

\newcommand{\TableFont}{\footnotesize}
\newcommand{\TableFontIt}{\ttit}
\newcommand{\SetTableFont}[1]{\renewcommand{\TableFont}{#1}}

\newcommand{\MakeTable}[4]{\begin{table}[htb]\TabCap{#2}{#3}
  \begin{center} \TableFont \begin{tabular}{#1} #4 
  \end{tabular}\end{center}\end{table}}

\newcommand{\MakeTableSep}[4]{\begin{table}[p]\TabCap{#2}{#3}
  \begin{center} \TableFont \begin{tabular}{#1} #4 
  \end{tabular}\end{center}\end{table}}

\newenvironment{references}%
{
\footnotesize \frenchspacing
\renewcommand{\thesection}{}
\renewcommand{\in}{{\rm in }}
\renewcommand{\AA}{Astron.\ Astrophys.}
\newcommand{\AAS}{Astron.~Astrophys.~Suppl.~Ser.}
\newcommand{\ApJ}{Astrophys.\ J.}
\newcommand{\ApJS}{Astrophys.\ J.~Suppl.~Ser.}
\newcommand{\ApJL}{Astrophys.\ J.~Letters}
\newcommand{\AJ}{Astron.\ J.}
\newcommand{\IBVS}{IBVS}
\newcommand{\PASP}{P.A.S.P.}
\newcommand{\Acta}{Acta Astron.}
\newcommand{\MNRAS}{MNRAS}
\renewcommand{\and}{{\rm and }}
\section{{\rm REFERENCES}}
\sloppy \hyphenpenalty10000
\begin{list}{}{\leftmargin1cm\listparindent-1cm
\itemindent\listparindent\parsep0pt\itemsep0pt}}%
{\end{list}\vspace{2mm}}

\def\TYLDA{~}
\newlength{\DW}
\settowidth{\DW}{0}
\newcommand{\dw}{\hspace{\DW}}

\newcommand{\refitem}[5]{\item[]{#1} #2%
\def\REFARG{#3}\ifx\REFARG\TYLDA\else, {\it#3}\fi
\def\REFARG{#4}\ifx\REFARG\TYLDA\else, {\bf#4}\fi
\def\REFARG{#5}\ifx\REFARG\TYLDA\else, {#5}\fi.}

\newcommand{\Section}[1]{\section{#1}}
\newcommand{\Subsection}[1]{\subsection{#1}}
\newcommand{\Acknow}[1]{\par\vspace{5mm}{\bf Acknowledgments.} #1}
\pagestyle{myheadings}

\def\thefootnote{\fnsymbol{footnote}}

\begin{center}
{\Large\bf The Optical Gravitational Lensing Experiment.\\
\vskip3pt
{\Large\bf {\it UBVI} Photometry of Stars\\}
\vskip3pt
{\Large\bf in Baade's Window}\footnote{Based on  observations obtained
with the 1.3~m Warsaw telescope at the Las Campanas  Observatory of the
Carnegie Institution of Washington.}}
\vskip1cm
{\bf
B.~~P~a~c~z~y~{\'n}~s~k~i$^1$,~~A.~~U~d~a~l~s~k~i$^2$,
~~M.~~S~z~y~m~a~{\'n}~s~k~i$^2$,~~M.~~K~u~b~i~a~k$^2$,
~~G.~~P~i~e~t~r~z~y~\'n~s~k~i$^2$,~~I.~~S~o~s~z~y~{\'n}~s~k~i$^2$,
~~P.~~W~o~\'z~n~i~a~k$^1$,~~ and~~K.~~\.Z~e~b~r~u~\'n$^2$}
\vskip3mm
{$^1$ Princeton University Observatory, Princeton, NJ 08544-1001, USA\\
e-mail: (bp,wozniak)@astro.princeton.edu\\
$^2$Warsaw University Observatory, Al.~Ujazdowskie~4, 00-478~Warszawa, Poland\\
e-mail: (udalski,msz,mk,pietrzyn,soszynsk,zebrun)@sirius.astrouw.edu.pl}
\end{center}

\Abstract{We present {\it UBVI} photometry for 8530 stars in Baade's
Window  obtained during the OGLE-II microlensing survey. Among these are
over one  thousand red clump giants. 1391 of them have photometry with
errors smaller than 0.04, 0.06, 0.12, and 0.20~mag in the {\it I, V, B,}
and {\it U}-band, respectively. We constructed a map of interstellar
reddening. The corrected colors of  the red clump giants: ${(U-B)_0}$,
${(B-V)_0}$, and ${(V-I)_0}$ are very well  correlated, indicating that
a single parameter determines the observed spread  of their values,
reaching almost 2~mag in the ${(U-B)_0}$. It seems most  likely that
heavy element content is the dominant parameter, but it is  possible
that another parameter: the age (or mass) of a star moves it along  the
same trajectory in the color--color diagram as the metallicity. The 
current ambiguity can be resolved with spectral analysis, and our
catalog may  be useful as a finding list of red clump giants. We point
out that these K  giants are more suitable for a fair determination of
the distribution of  metallicity than brighter M giants. 

We also present a compilation of {\it UBVI} data for 308 red clump
giants near  the Sun, for which Hipparcos parallaxes are more accurate
than 10\%. Spectral  analysis of their metallicity may provide
information about the local  metallicity distribution as well as the
extent to which mass (age) of these  stars affects their colors. 

It is remarkable that in spite of a number of problems, stellar models
agree  with observations at the 0.1--0.2~mag level, making red clump
giants not only  the best calibrated but also the best understood
standard candle.}{~} 

\Section{Introduction}
Red clump giants appear to be among the best standard candles
available. Not  a single RR~Lyr star or a Cepheid has its parallax
measured by Hipparcos  (Perryman \etal 1997) with a 10\% accuracy
(Horner \etal 1999), but there are  about $10^3$ red clump giants with
parallaxes at least as good as 10\%  (Paczy\'nski and Stanek 1998). The
absolute {\it I}-band magnitude of these  stars appears to be almost
independent of their ${(V-I)}$ color (Paczy\'nski  and Stanek 1998). The
weak dependence on metallicity was empirically  calibrated by Udalski
(1998a), who also demonstrated that there is no  detectable dependence
on age, as long as it is between $2\times10^9$ and  $10\times10^9$~years
(Udalski 1998b). While this calibration is not perfect,  it is superior
to the calibration of either RR~Lyr or Cepheid variables, or  any other
standard candle ever proposed. Unfortunately, there is some  confusion
about the color--metallicity dependence (Paczy\'nski 1998). 

The main purpose of this paper is to present a catalog of {\it UBVI} 
photometry for 8530 stars in Baade's Window obtained during the second
phase  of the Optical Gravitational Lensing Experiment (OGLE-II)
microlensing project  (\cf Udalski, Kubiak and Szyma\'nski 1997). The
important part of the catalog  is the list of 1391 red clump giants in
the galactic bulge/bar, covering a very  broad range of colors, and
presumably metallicities. We also present a  compilation of {\it UBVI}
photometry (Mermilliod and Mermilliod 1994) for 308  nearby red clump
giants, for which Hipparcos parallaxes are more accurate than  10\%. Our
expectation is that these two data sets provide a convenient list of 
objects for which metallicity can be obtained spectroscopically in order
to  determine the range of [Fe/H] and [O/H] for nearby stars as well as
for those  near the Galactic center. Such spectroscopic study could
resolve the current  ambiguity about the distribution of metallicities
and the relation between the  metallicity and colors of red clump
giants. 

We also present, following Kiraga, Paczy\'nski and Stanek (1997), a 
description of the procedure which appears to be fairly accurate in
mapping  interstellar extinction. The variation, with the line of sight,
of average  colors of the bulge red clump stars, red sub-giants, and the
stars near the  main sequence turn off point are very well correlated
with each other, as all  are affected by the same variations in the
interstellar reddening. The map of  color variations provides
information about angular scale over which the  extinction varies. 

\Section{Observations}
Observations presented in this paper were collected during the second
phase of  the OGLE microlensing search with the 1.3-m Warsaw telescope
at the Las  Campanas Observatory, Chile, which is operated by the
Carnegie Institution of  Washington. The telescope was equipped with the
"first generation" camera with  a SITe ${2048\times2048}$ CCD detector.
The pixel size was 24~$\mu$m giving  the 0.417~arcsec/pixel scale.
Observations were performed in the "medium"  speed reading mode of CCD
detector with the gain 7.1~e$^-$/ADU and readout  noise of about
6.3~e$^-$. Details of the instrumentation setup can be found in 
Udalski, Kubiak and Szyma{\'n}ski (1997). 

The {\it BVI} photometry comes from observations of the OGLE-II field
covering  large part of Baade's Window: BUL$\_$SC45. Observations of
this field were  obtained in the drift-scan mode of CCD detector and
the field covers  approximately ${14.2\times57}$~arcmin in the sky.
Coordinates of its center  are: ${\rm
RA(J2000)=18\uph03\upm37}$, ${\rm
DEC(J2000)=-30\arcd05\arcm00\arcs}$. The effective exposure time was 87,
124 and 162 seconds for the {\it I, V} and {\it B}-band, respectively.
BUL$\_$SC45 field  is observed in somewhat different way than the
remaining 47 bulge fields -- it  is not a subject of microlensing
search. Observations of this field are  collected mainly for maintaining
the phasing of variable stars discovered in  its part during the first
stage of the OGLE project (\cf Udalski \etal 1995).  Therefore only
about 30 {\it I}-band, 15 {\it V}-band and 10 {\it B}-band  epochs were
collected during the period August~5, 1997 through October~21,  1998. 

The {\it U}-band photometry was obtained in the standard, still frame
mode of  the CCD detector. One field, covering ${14.2\times14.2}$~arcmin
in the sky and  centered approximately at the field BWC of OGLE-I: ${\rm
RA(J2000)=18\uph03\upm24\ups}$, ${\rm
DEC(J2000)=-30\arcd02\arcm00\arcs}$;  (\cf Udalski \etal 1993, and also
Stanek 1996) was observed on four nights  between September~20 and
September~25, 1998. The field overlaps in large part  with BUL$\_$SC45
field. The exposure time was 1800 seconds. 

Observations collected in the drift-scan mode were reduced with the
standard  OGLE data pipeline and transformed to the {\it BVI} system
based on  observations of standard stars from Landolt (1992) fields
collected during  5--7 photometric nights (\cf Udalski \etal 1998b).
{\it U}-band photometry  collected in the still-frame mode was reduced
in similar manner as described  in Udalski (1998b) with the difference
that the full ${2048\times2048}$ pixel  image was divided into 16
${512\times512}$ pixel sub-frames and each of them  was reduced
separately to ensure constant PSF. In this case transformation to  the
standard system was based on photometry of standard stars collected
during  three photometric nights. 

The systematic error of the {\it BVI}-band transformation should be
smaller  than 0.02--0.03~mag. For {\it U}-band the systematic error
might be larger and  we conservatively assume it to be equal to
0.05~mag. This is caused by  somewhat different spectral response of the
OGLE instrumental ultraviolet  filter because of somewhat steeper cut of
the short wavelength  (${\lambda<3500}$~\AA) side of the {\it U}-band by
the telescope field  corrector made from BK7 Schott glass. Nevertheless,
the standard stars  transform well to the standard system and the color
term coefficient for the {\it U}-band is  equal to 0.127 (0.000 means
the standard system) which compares to $-0.041$,  $+0.004$ and $+0.032$
for {\it B, V} and {\it I}-band, respectively. 

The final list of Baade's Window stars with {\it UBVI} photometry was 
constructed by cross-identification of all stars found in the {\it
U}-band  images with the stars from the drift-scan field BUL$\_$SC45
using second order  transformation between pixel coordinates. 

Astrometric positions of stars were determined with the standard OGLE 
procedure as described in Udalski \etal (1998b). About 10270 stars from 
the BUL$\_$SC45 field were used to derive transformation between image pixel
coordinates and the Digitized Sky Survey (DSS) coordinate system.
Internal  accuracy of determined equatorial coordinates is about
0.15~arcsec with  possible systematic errors of the DSS system up to
0.7~arcsec.

\Section{{\it UBVI} Photometry of Baade's Window Stars}
Table~1 contains {\it UBVI} photometric data for 8530 stars in Baade's
Window.  We provide there star ID number (in BUL$\_$SC45 field),
equatorial coordinates  (J2000) and observed and extinction-free (see
below) {\it UBVI} magnitudes of  each star. Table~1 is too large to be
conveniently printed, therefore only a  few sample lines are shown. Full
versions of Tables presented in this paper  are available in electronic
form from the OGLE Internet data archive:  {\it
http://www.astrouw.edu.pl/\~{}ftp/ogle} or its mirror\newline
{\it http://www.astro.princeton.edu/\~{}ogle}. 
\begin{figure}[htb]
\vspace*{-5mm}
\psfig{figure=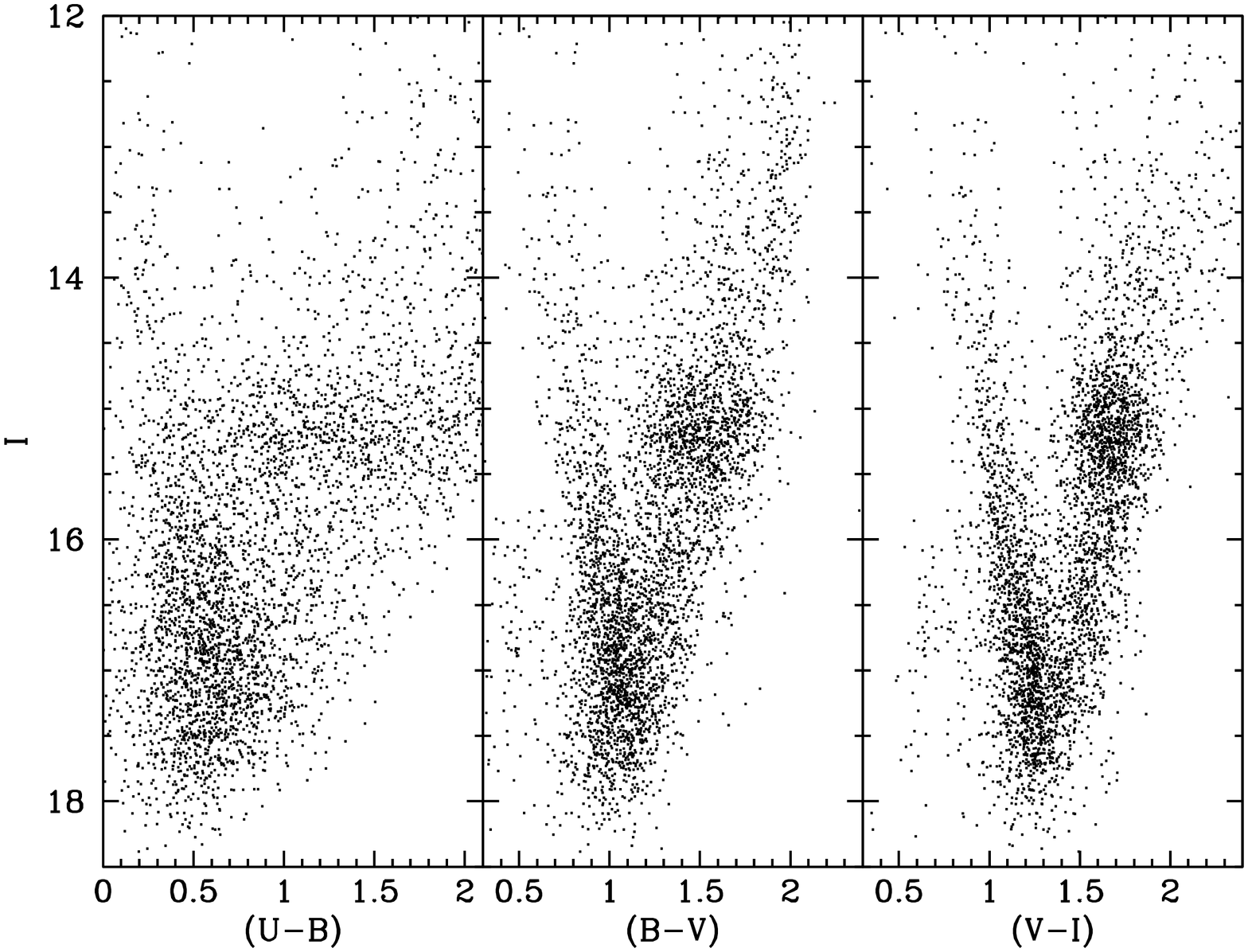,bbllx=0pt,bblly=0pt,bburx=590pt,bbury=750pt,width=12cm,clip=,angle=-90}
\FigCap{Color--magnitude diagrams for 50\% of the 8530 stars in Baade's Window 
for which {\it UBVI} photometry was obtained by the OGLE-II. Photometric 
errors in all bands are smaller than 0.1 mag. Notice the increase of color 
range of red clump giants progressing from $(V-I)$, through $(B-V)$ to $(U-
B)$.} 
\end{figure}

\begin{figure}[p]
\vspace*{2cm}
\centerline{\hglue8.3cm\psfig{figure=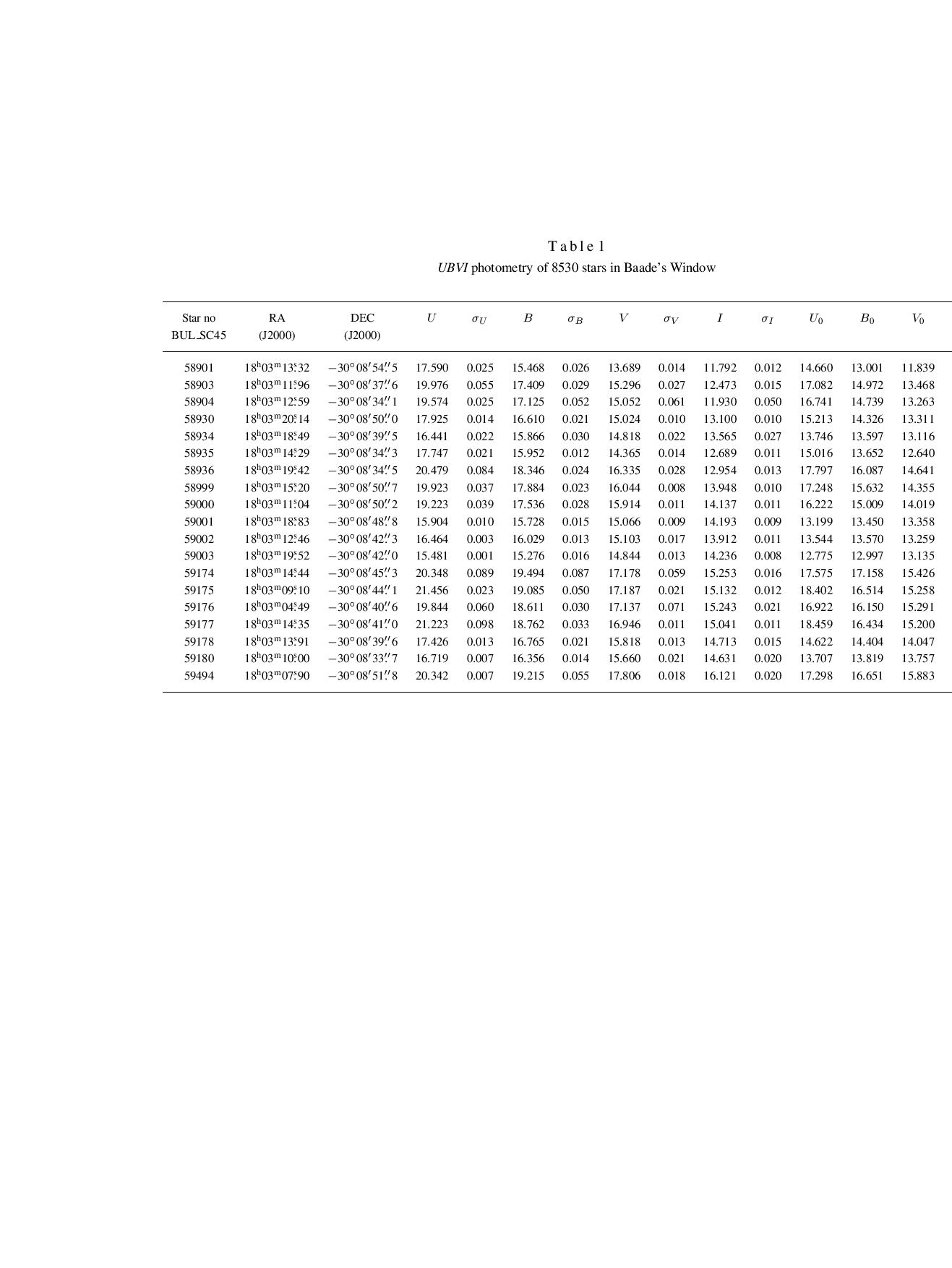,height=23cm,angle=90}}
\end{figure}
The ${I-(V-I)}$, ${I-(B-V)}$, and ${I-(U-B)}$ color--magnitude diagrams
(CMDs)  of Baade's Window stars are shown in Fig.~1 for 50\% of the
total of 8530  stars for which {\it UBVI} band photometry was more
accurate than 0.1~mag. The  red clump is clearly visible in all three
diagrams, and its width increases  dramatically toward the blue and
ultraviolet. For these diagrams to be useful  we have to correct them
for interstellar reddening. 

\Section{Reddening Map}
We followed Kiraga, Paczy\'nski and Stanek (1997) in the determination
of  interstellar reddening variation over our field. We selected 38,249
stars for  which {\it V} and {\it I}-band photometry was more accurate
than 0.1~mag. The  corresponding CMD for 25\% of those stars is shown in
Fig.~2. Four regions  were selected: one near the red clump (RC), one
below it, at the red giant  branch (RG), and two near the main sequence
turn off point, one brighter, and  the other one fainter (TOPb, TOPf),
as indicated in Fig.~2. The six parallel  lines are described with the
equations: ${I-1.5\times(V-I)=C}$, where   ${C=12.00}$, 13.25, 14.50,
15.50, 16.00, 16.50, respectively. The slope of  these lines was chosen
so that they are parallel to the reddening vector  (Stanek 1996). 
\begin{figure}[htb]
\hglue1.3cm\psfig{figure=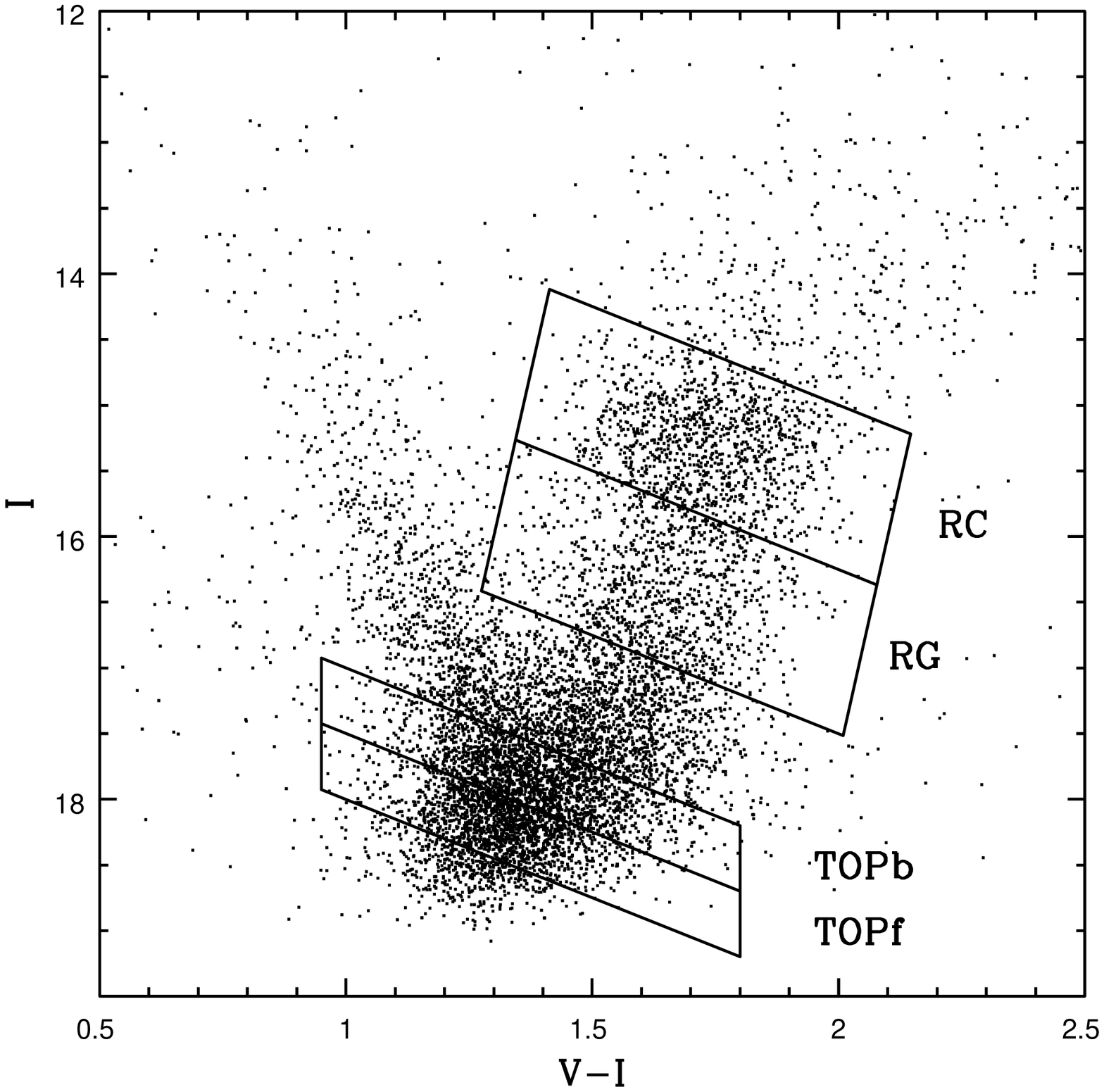,bbllx=20pt,bblly=150pt,bburx=550pt,bbury=680pt,width=10cm,clip=}
\vspace*{3pt}
\FigCap{Color--magnitude diagram for 25\% of the 38,249 stars in Baade's 
Window for which {\it V} and {\it I}-band photometry was obtained by the 
OGLE-II. Photometric errors in both bands are smaller than 0.1~mag. The stars 
in the four marked regions were used for reddening determination.} 
\end{figure}

We divided our field into squares of ${200\times200}$ pixels. In every
square  we calculated average color of the stars in each of the four
groups  separately, and also the average of the stars in all four
groups. It is safe  to assume that there was no significant variation in
stellar populations  within our field, which had the total area of only
$(0\zdot\arcd23)^2$. The  variation of average colors was due to
variation in the interstellar reddening  as well as the statistical
fluctuations caused by a small number of stars. 

The colors of all four stellar groups are very well correlated, as shown
in  Fig.~3. To bring the colors of the four groups to the same average
value the  corrections were added to the ${(V-I)}$ color of each star,
with ${\Delta(V- I)_{\rm TOPf}=+0.200}$~mag, ${\Delta(V-I)_{\rm
TOPb}=+0.138}$~mag, $\Delta(V-I)_{\rm RG}=-0.120$~mag and
$\Delta(V-I)_{\rm RC}=-0.217$~mag. The shift  between colors of each
group in Fig.~3 has been increased for the clarity of  presentation. 
\begin{figure}[htb]
\hglue1.5cm\psfig{figure=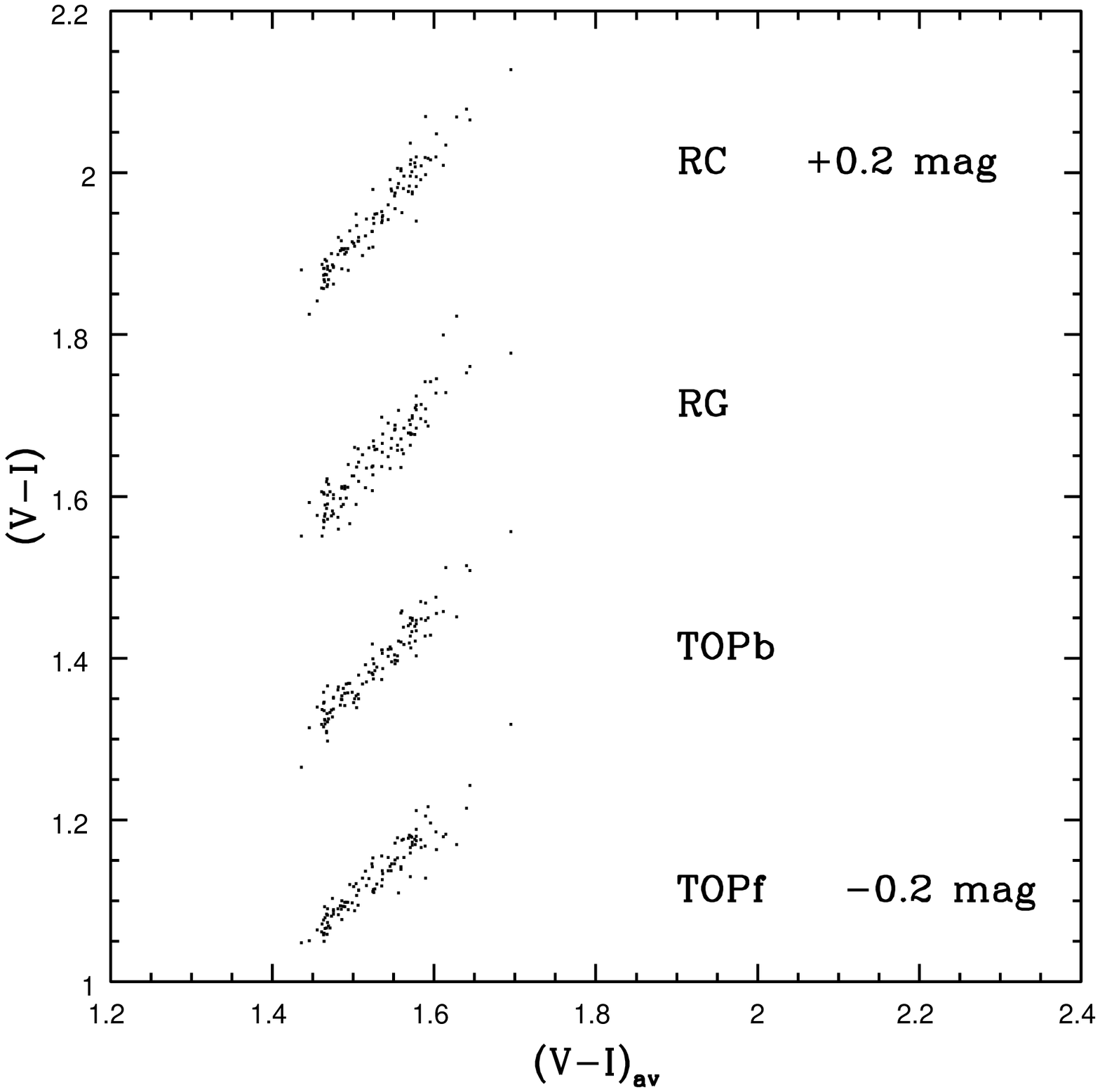,bbllx=20pt,bblly=145pt,bburx=550pt,bbury=680pt,width=9cm,clip=}
\vspace*{3pt}
\FigCap{Relation between the average ${(V-I)}$ colors of the stars in four 
areas indicated in Fig.~2 are shown as a function of the average $(V-I)_{\rm 
av}$ for stars in all four areas. Each dot corresponds to a different square 
in the CCD field, $200\times 200$ pixels, \ie ${83\zdot\arcs4\times 
83\zdot\arcs4}$ each.}
\end{figure} 
\begin{figure}[htb]
\hglue1.5cm\psfig{figure=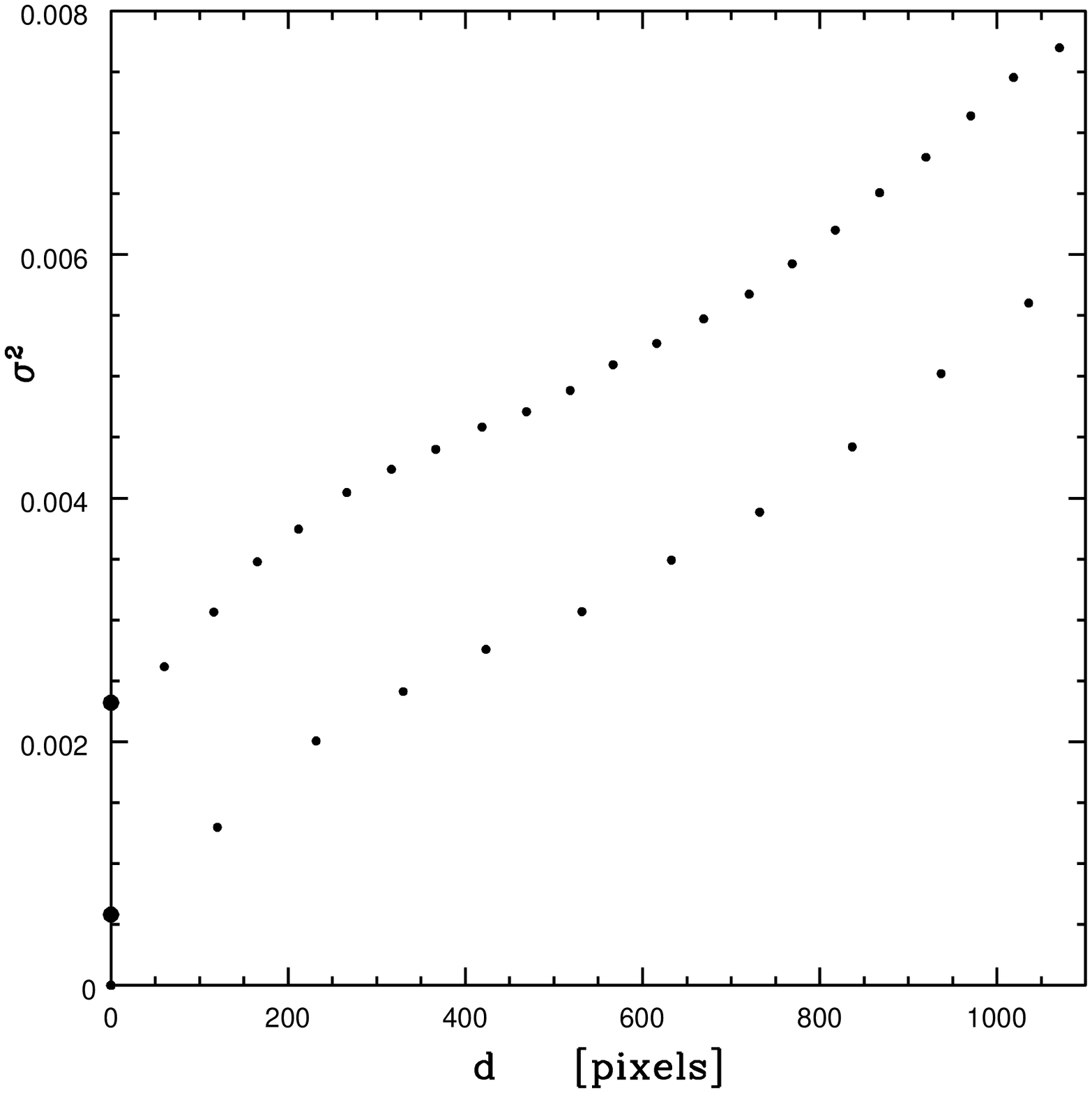,bbllx=20pt,bblly=145pt,bburx=550pt,bbury=680pt,width=9cm,clip=}
\vspace*{3pt}
\FigCap{Correlation function $\sigma^2$ of the variation in ${(V-
I)}$ is shown as a function of separation~$d$. The upper sequence of points 
corresponds to the resolution ${50\times50}$ pixels, while the lower sequence 
corresponds to the resolution ${100\times100}$ pixels.} 
\end{figure}

As there was such a good correlation between the colors of the four
groups we  combined all stars, adding the suitable corrections as given
above. We  calculated average colors for the stars in squares
${50\times50}$ pixels, \ie   ${20\zdot\arcs8\times20\zdot\arcs8}$ in the
sky. The total number of such  squares in our field was
${39\times42=1638}$. These became our resolution  elements for the
reddening map, or the elementary squares. The number of stars  in the
four groups was: ${N_{\rm TOPf}=6422}$, ${N_{\rm TOPb}=7776}$, ${N_{\rm 
RG}=4185}$, ${N_{\rm RC}=4546}$. The total number of all stars used for
the  reddening determination was ${N=22,929}$, the average number of
stars per  elementary square was ${n=14.0}$, and the variance of the
average color was  ${\langle\sigma^2_{V-I}\rangle=(0.034~{\rm mag})^2}$.
The {\it rms} variation  of the ${(V-I)}$ color per star was
approximately 0.13~mag. 

Given the values of the average ${(V-I)}$ colors in 1638 elementary
squares we  calculated the correlation function defined as
$$\sigma^2(d)=\langle\Delta(V- I)^2\rangle,\eqno(1)$$ averaged over all
pairs of squares within each range of  separations~$d$. $\sigma^2$ is
shown in Fig.~4 as a function of the separation  between the elementary
squares (the upper sequence of points). The point  corresponding to the
separation ${d=0}$ is twice the variance of the ${(V-I)}$  color per
elementary square. Fig.~4 shows that the value of the correlation 
function doubles at a separation ${d\approx500}$ pixels, which is 10
times  larger than the elementary square size. We repeated the exercise
by combining  4 adjacent elementary squares into one composite square of
${100\times100}$  pixels, thereby increasing the average number of stars
to 56.0, and reducing  the variance of the average ${(V-I)}$ color to
$(0.017~{\rm mag})^2$. The  corresponding correlation points are shown
as the lower sequence in Fig.~4.  This time the correlation function
doubles its value at a separation  ${d\approx 100}$ pixels. 

While the zero point depends on the choice of the square size, the rest
of the  correlation function is the same in both cases, except for the
zero point  shift. The contribution of interstellar reddening to the
correlation function  may be approximated as 
$$\langle\Delta E_{V-I}^2\rangle\approx\left(0.027~{\rm mag}\right)^2\times
{d\over1'},\qquad 0<d<7'.\eqno(2)$$

How to chose the right square size? If the squares are small then the 
resolution is high, and the variations of interstellar reddening from
square  to square are small. However, there are few stars per square and
the  statistical error in the estimate of the average stellar color is
large. When  the squares are large the opposite is true: the average
color is calculated  accurately, but the resolution is poor and the
variations of the interstellar  reddening within the square become
substantial. Without attempting a rigorous  analysis it seems that a
sensible square size is such, that the two  contributions: the random
noise due to a finite number of stars within the  square, and the
variations of the interstellar reddening within the square are 
comparable. Fig.~4 suggests that a sensible choice is a square size
${100\times  100}$~pixels, \ie ${41\zdot\arcs7\times41\zdot\arcs7}$ in
the sky. 
\begin{figure}[htb]
\hglue1.5cm\psfig{figure=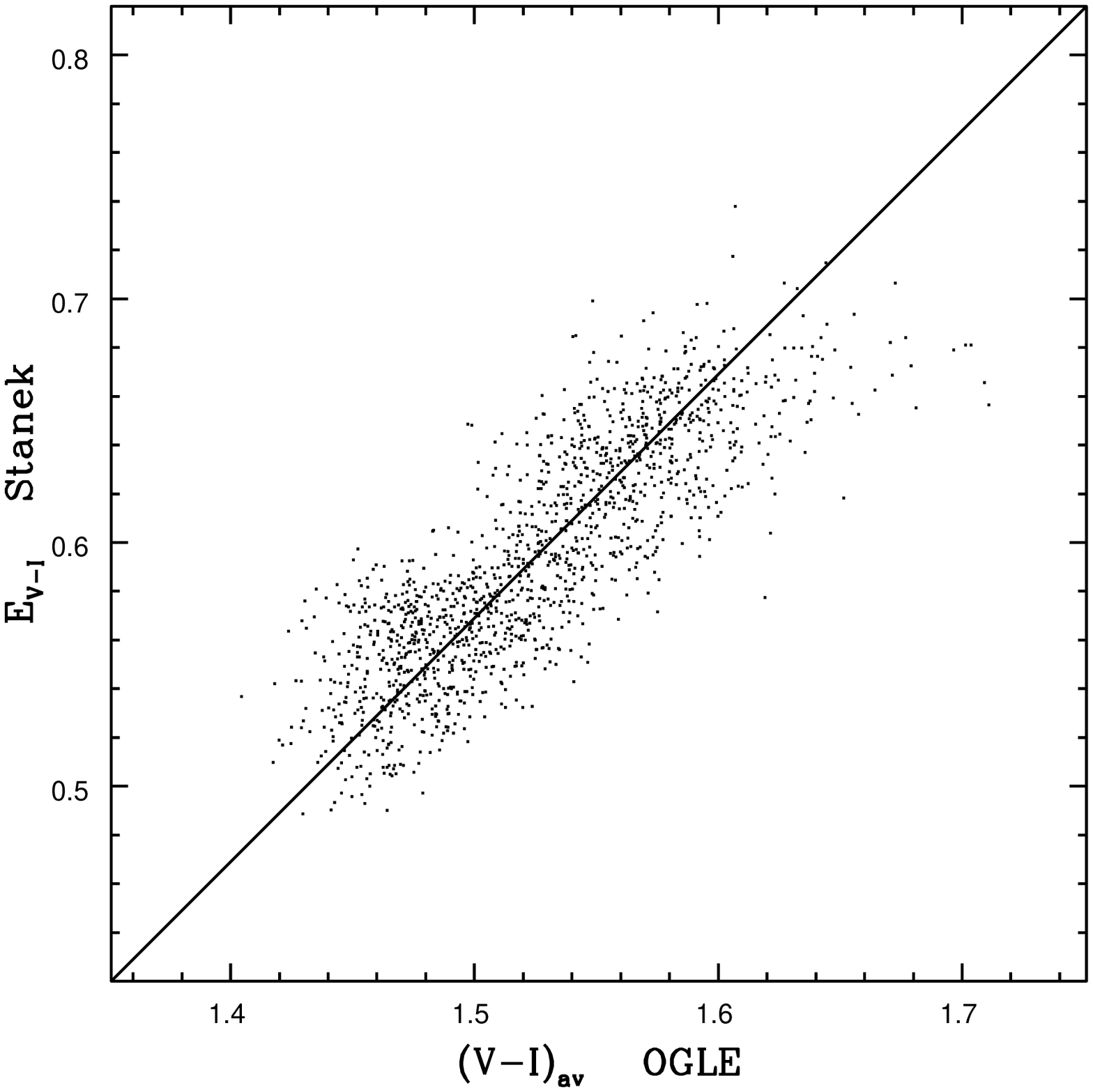,bbllx=20pt,bblly=145pt,bburx=550pt,bbury=680pt,width=9cm,clip=}
\vspace*{3pt}
\FigCap{Correlation of the average color ${(V-I)_{\rm av}}$ of the OGLE 
stars on the reddening ${E_{V-I}}$ calculated with Stanek's (1996) map. The 
diagonal line is the relation: ${(V-I)_{\rm av}=E_{V-I}+0.931}$.} 
\end{figure}

We calculated average ${(V-I)_{\rm av}}$ color for the stars in a grid
of  ${100\times100}$ pixel squares. We shifted the grid by 50 pixels in
both  directions, so as to have average color calculated at 50 pixel
intervals. A  comparison between the ${(V-I)_{\rm av}}$ color from our
data and the color  excess ${E_{V-I}}$ calculated with the Stanek's
(1996) extinction map is shown  in Fig.~5. The agreement is good, and we
adopt the relation: 
$$E_{V-I}=(V-I)_{\rm av}-0.931,\qquad\langle E_{V-I}\rangle=0.59.\eqno(3)$$ 
Obviously, our method, as well as Stanek's, provides only a differential
reddening map. The zero point adopted by Stanek, and therefore also by
us, is  based on the determination by Gould \etal (1998) and Alcock
\etal (1998). 

We calculate reddening within our $(0\zdot\arcd23)^2$ field by linear 
interpolation in the grid with the spacing of 50 pixels. The values of
$E_{V-I}$ are based on the average ${(V-I)_{\rm av}}$ calculated in
${100\times100}$  pixel squares. Therefore, the resolution of our
extinction map is 100~pixels,  \ie 41\zdot\arcs7 in the sky. We estimate
the statistical accuracy of our  reddening map to be
${\approx0.02}$~mag. Of course, the uncertainty in the zero  point of
our extinction is substantially larger. The {\sc FORTRAN} code {\sc 
red.f}, calculating reddening within our field is available from the
OGLE Internet archive. 

\Section{Color--Color Diagrams}
The extinction in the four bands was calculated according to the formulae:
$$A_I=1.500\cdot E_{V-I},\qquad\langle A_I\rangle=0.88,\eqno{\rm (4a)}$$
$$A_V=2.500\cdot E_{V-I},\qquad\langle A_V\rangle=1.48,\eqno{\rm (4b)}$$
$$A_B=3.333\cdot E_{V-I},\qquad\langle A_B\rangle=1.97,\eqno{\rm (4c)}$$
$$A_U=3.958\cdot E_{V-I},\qquad\langle A_U\rangle=2.34,\eqno{\rm (4d)}$$
and the colors were corrected for the reddening accordingly. The coefficients 
in the Eqs.~(4a)--(4d) are more or less standard, but the reader may adopt 
different values as we provide the original as well as the dereddened 
photometry. The reddening varied relatively little over our small field of 
view, with 90\% of stars in the range ${0.51<E_{V-I}<0.65}$. Therefore, the 
CMDs corrected for extinction were similar to those shown in Fig.~1, except 
for a shift in color and magnitude. 

\begin{figure}[p]
\vspace*{2cm}
\centerline{\hglue8.3cm\psfig{figure=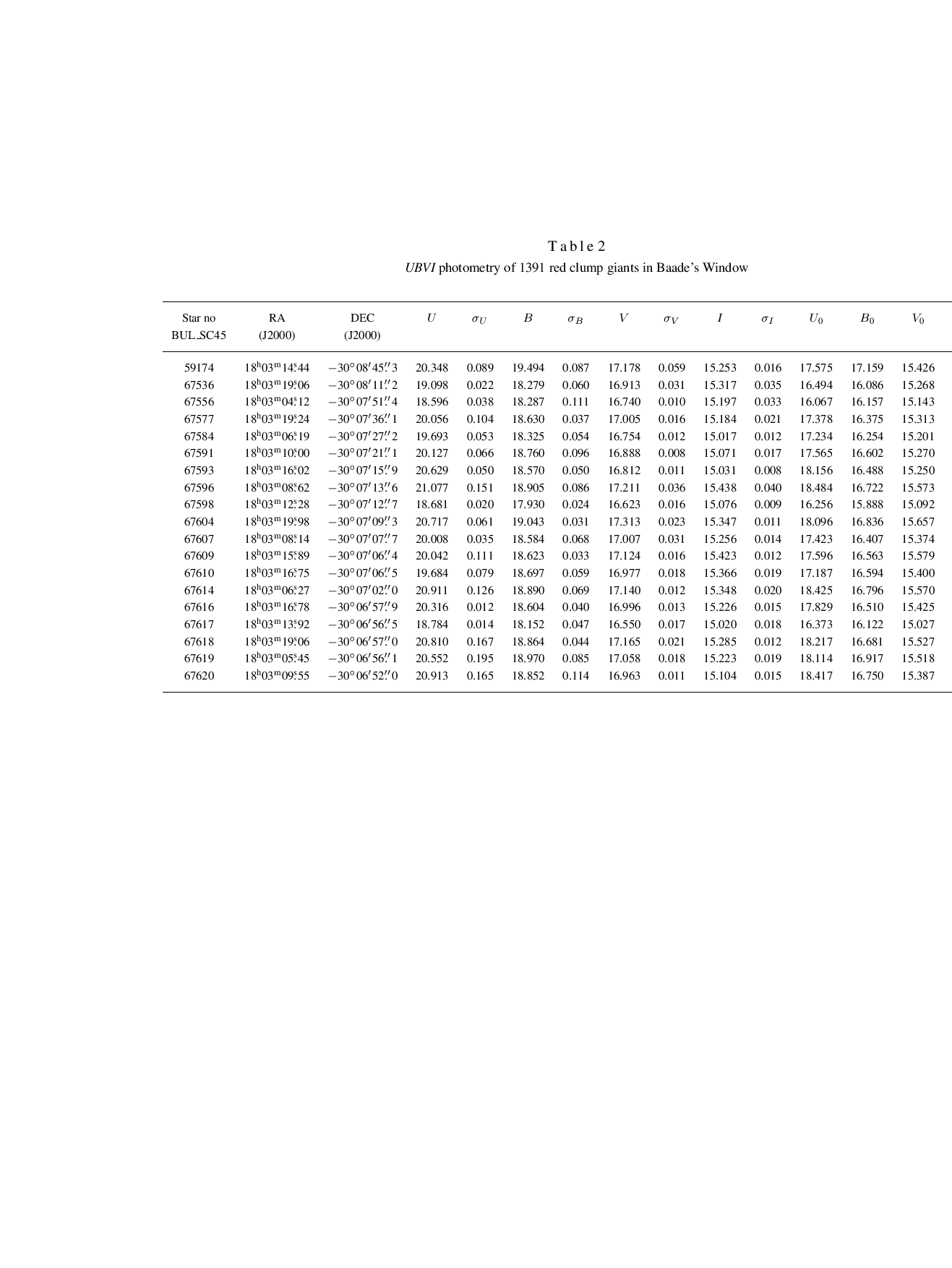,height=23cm,angle=90}}
\end{figure}
\begin{figure}[p]
\vspace*{2cm}
\centerline{\hglue8.3cm\psfig{figure=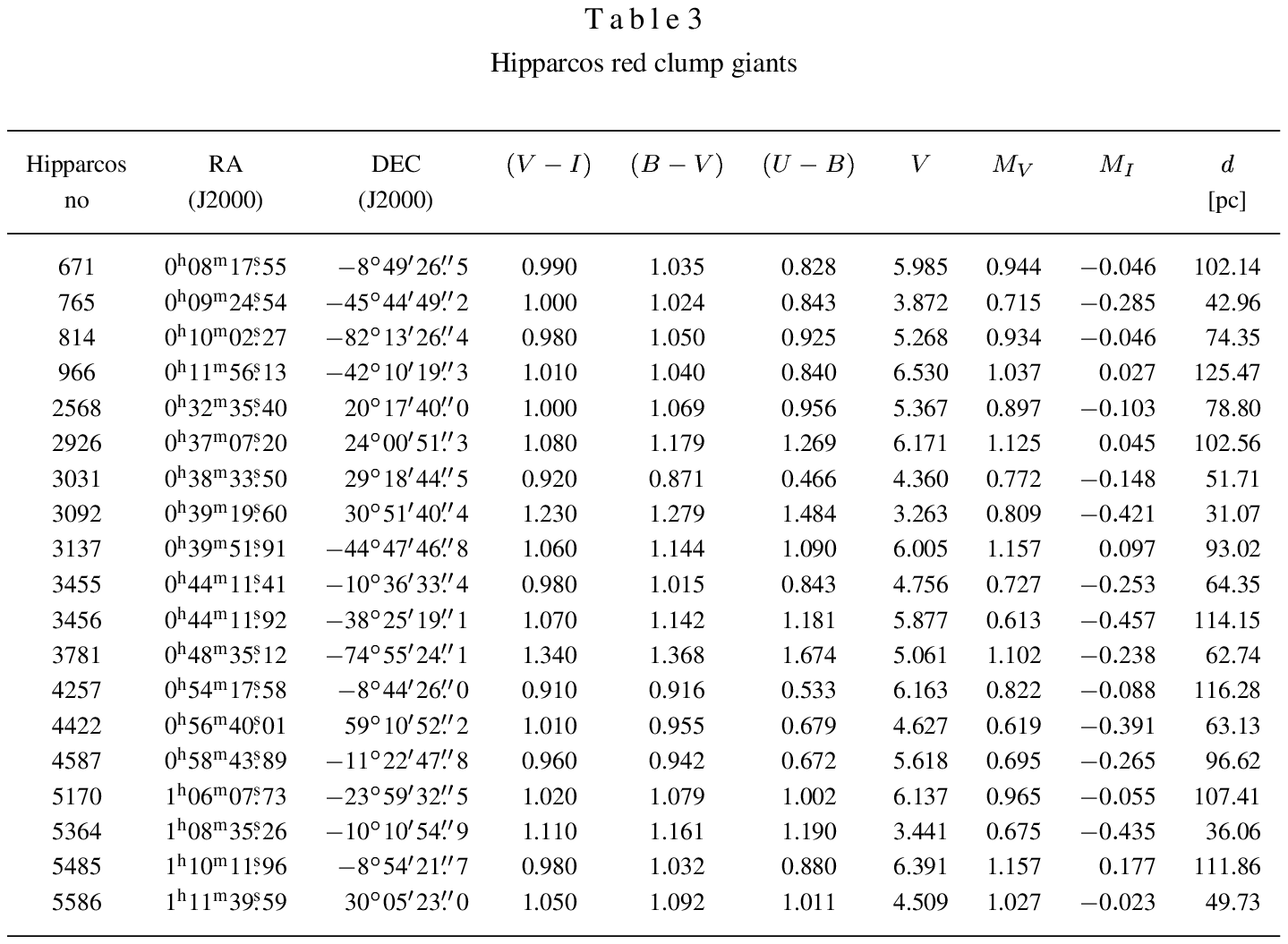,height=23cm,angle=90}}
\end{figure}

Median errors of the OGLE-II photometry of the red clump giants are 
0.016, 0.02,  0.04 and 0.06~mag for {\it I, V, B} and {\it U}-band,
respectively. The {\it I}-band  errors are color independent, but the
errors in {\it V, B} and mostly in {\it U}-band increase with the
${(V-I)}$ color, because the redder stars are fainter. In the {\it
U}-band the reddest red clump stars are near our detection limit, as it
is apparent in Fig.~1. We rejected stars with the errors larger than
0.04, 0.06, 0.12, and 0.20~mag in the {\it I, V, B} and {\it U}-band,
respectively; these were about 5\% of the total. 1391 stars had smaller
photometric errors and satisfied the inequalities:
$$14.0<I_0<14.5,\qquad0.8<(V-I)_0<1.6,\eqno(5)$$
and these are listed in Table~2 (full version is available from the OGLE
Internet archive). Note, that some stars in Table~2 are not present in
Table~1, because the criteria used for the error discrimination were
different for the two sets.

We also selected red clump giants from the Hipparcos catalog according
to the following criteria: genuine {\it I}-band photometry was available
in the catalog, the absolute magnitude $M_I$, the color ${(V-I)}$, and the
parallax $\pi$ satisfied the criteria: 
$$-0.5<M_I<0.2,\qquad0.8<(V-I)<1.6,\qquad\pi~{\rm error<10\%},\eqno(6)$$ 

J-C. Mermilliod kindly provided us with the {\it UBV} photometry for the
selected stars in a computer readable form (\cf Mermilliod and
Mermilliod  1994, and {\it http://obswww.unige.ch/gcpd/} ). The list of
308 stars which  satisfied the criteria given above, had {\it UBVI}
photometry and had the  differences in {\it V} and {\it B} magnitudes as
given by the Hipparcos/Tycho  catalogs and by Mermilliod not larger than
0.03~mag, are listed in Table~3  (full version is available from the
OGLE Internet archive). As all Hipparcos stars with accurate parallaxes
are  within about 100~pc, their reddening is very small (Stanek and Garnavich
1998,  Sekiguchi and Fukugita 1999). 

The number of red clump giants for which Hipparcos provides accurate 
parallaxes is several times larger than 308, and they all have accurate
{\it  V} and {\it B} photometry, but a large fraction does not have
either {\it I}  or {\it U} data. The calibration of red clump giants
could be improved  considerably if standard {\it I} and {\it U}
photometry were made for these  bright stars. 

Figs.~6 and 7 show ${(V-I)_0-(B-V)_0}$ and ${(B-V)_0-(U-B)_0}$
color--color diagrams constructed for 1391 Baade's Window red clump
giants. Figs.~8 and 9 show similar diagrams for the Hipparcos sample of
red clump stars. The arrow  pointing the direction of interstellar
reddening is shown in Figs.~6 and 7, with the filled circles indicating
the average reddening, and the  open circles indicating the full range
of reddening within our field.
\begin{figure}[p]
\hglue2cm\psfig{figure=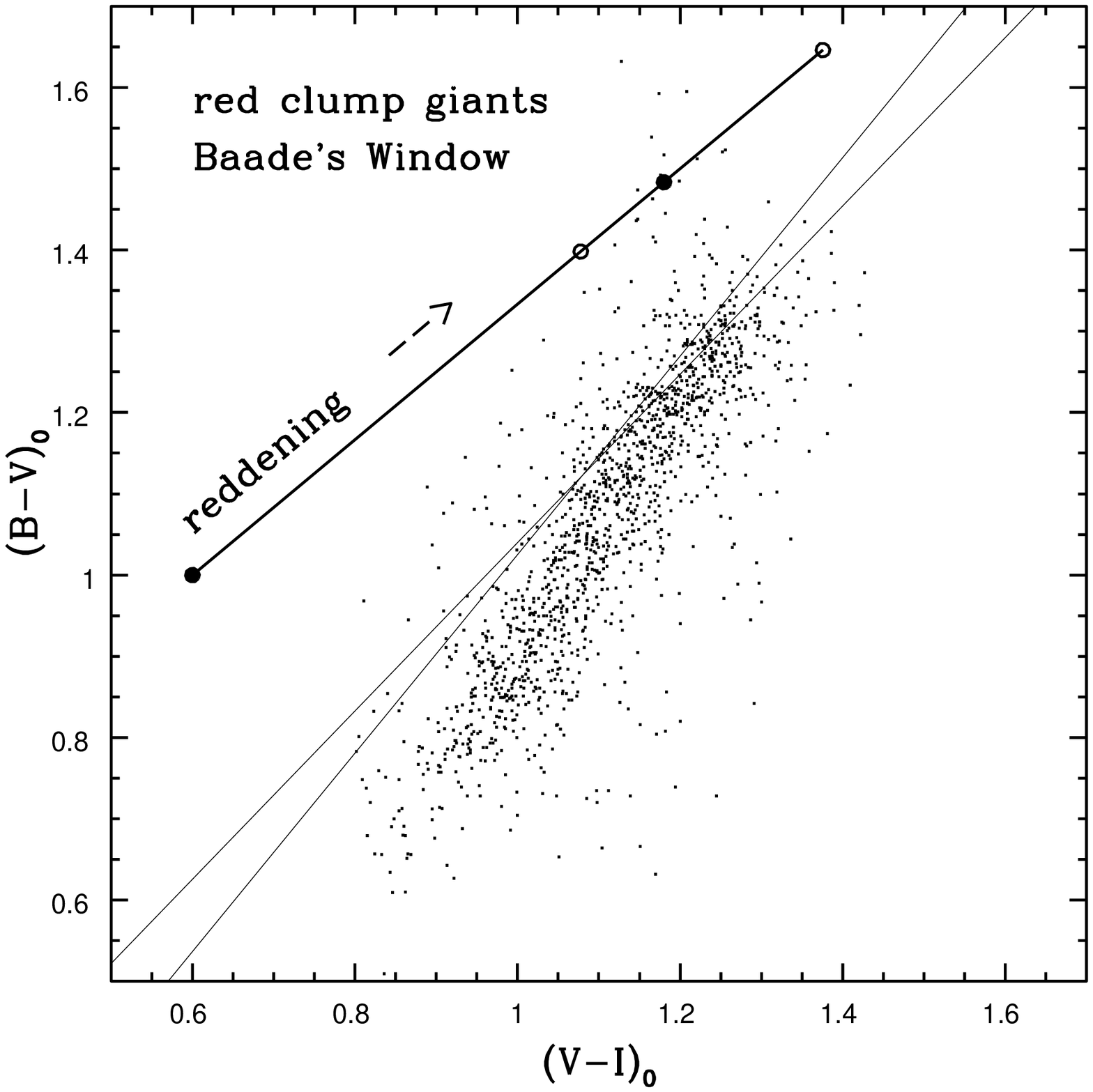,bbllx=20pt,bblly=145pt,bburx=550pt,bbury=680pt,width=8cm,clip=}
\vspace*{3pt}
\FigCap{Color--color diagram: ${(V-I)_0-(B-V)_0}$ for 1391 red clump
giants  in Baade's Window corrected for interstellar reddening. The two
thin lines that cross are for  reference; they are the same as in
Fig.~8.} 
\vspace*{10pt}
\hglue2cm\psfig{figure=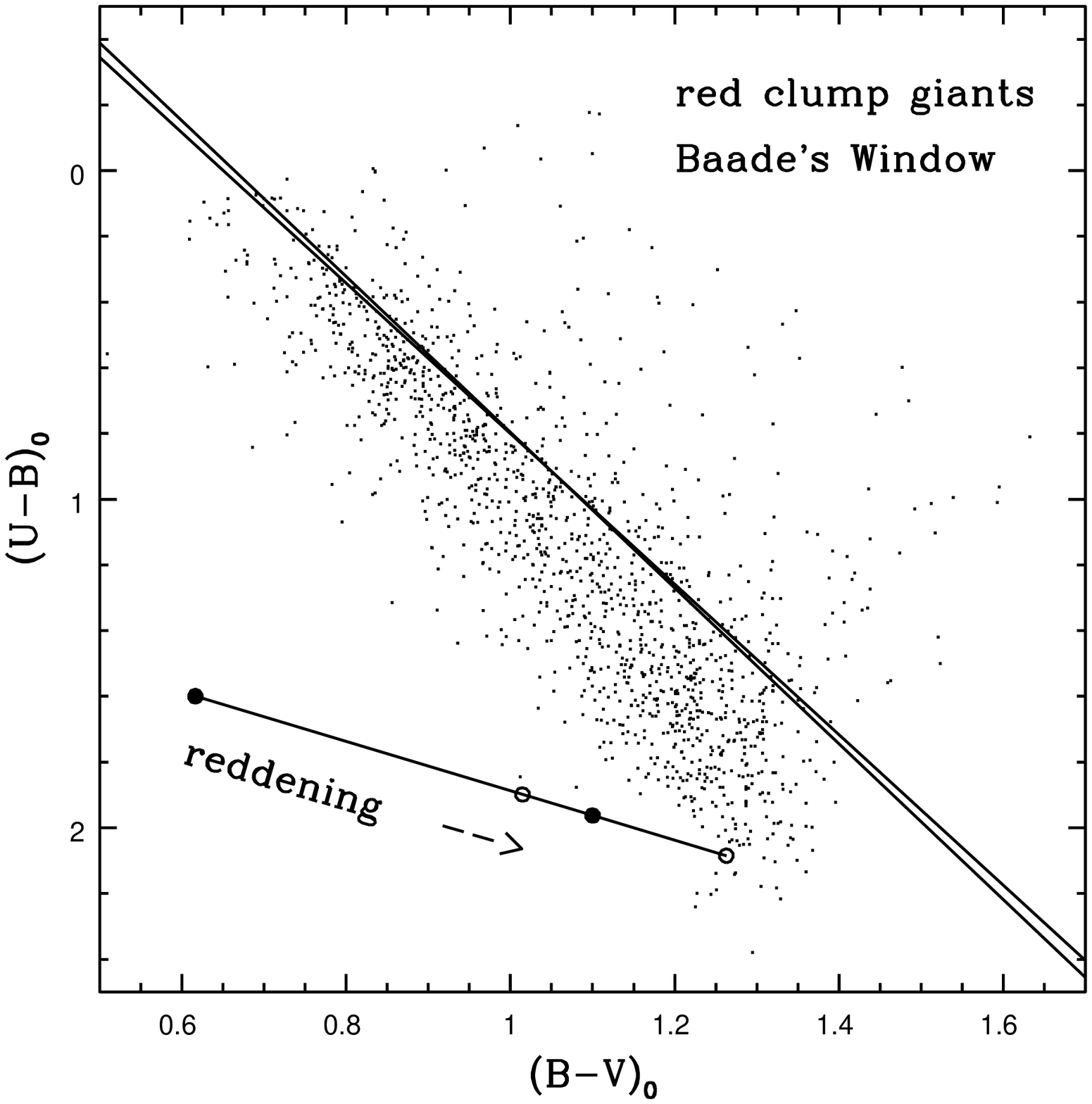,bbllx=20pt,bblly=145pt,bburx=550pt,bbury=680pt,width=8cm,clip=}
\vspace*{3pt}
\FigCap{Color--color diagram: ${(B-V)_0-(U-B)_0}$ for 1391 red clump
giants in Baade's Window corrected for interstellar reddening. The two
thin lines that cross are for  reference; they are the same as in
Fig.~9.}  \end{figure}

\begin{figure}[p]
\hglue2cm\psfig{figure=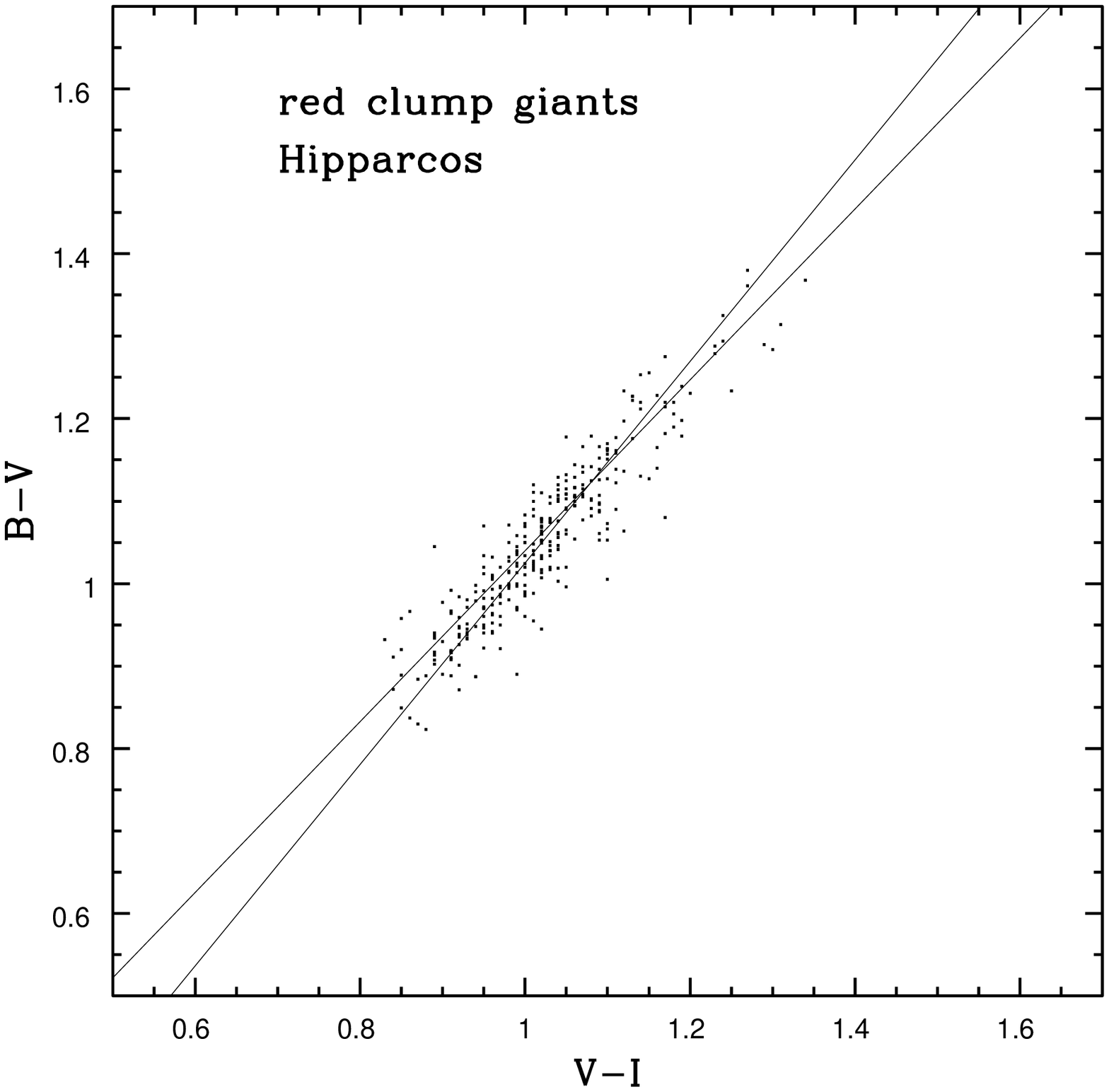,bbllx=20pt,bblly=145pt,bburx=550pt,bbury=680pt,width=8cm,clip=}
\vspace*{3pt}
\FigCap{Color--color diagram: ${(V-I)-(B-V)}$ for 308 red clump giants for 
which Hipparcos measured parallaxes with the accuracy better than 10\%. The 
two thin lines that cross are the regression lines of one color with respect 
to another.} 
\vspace*{10pt}
\hglue2cm\psfig{figure=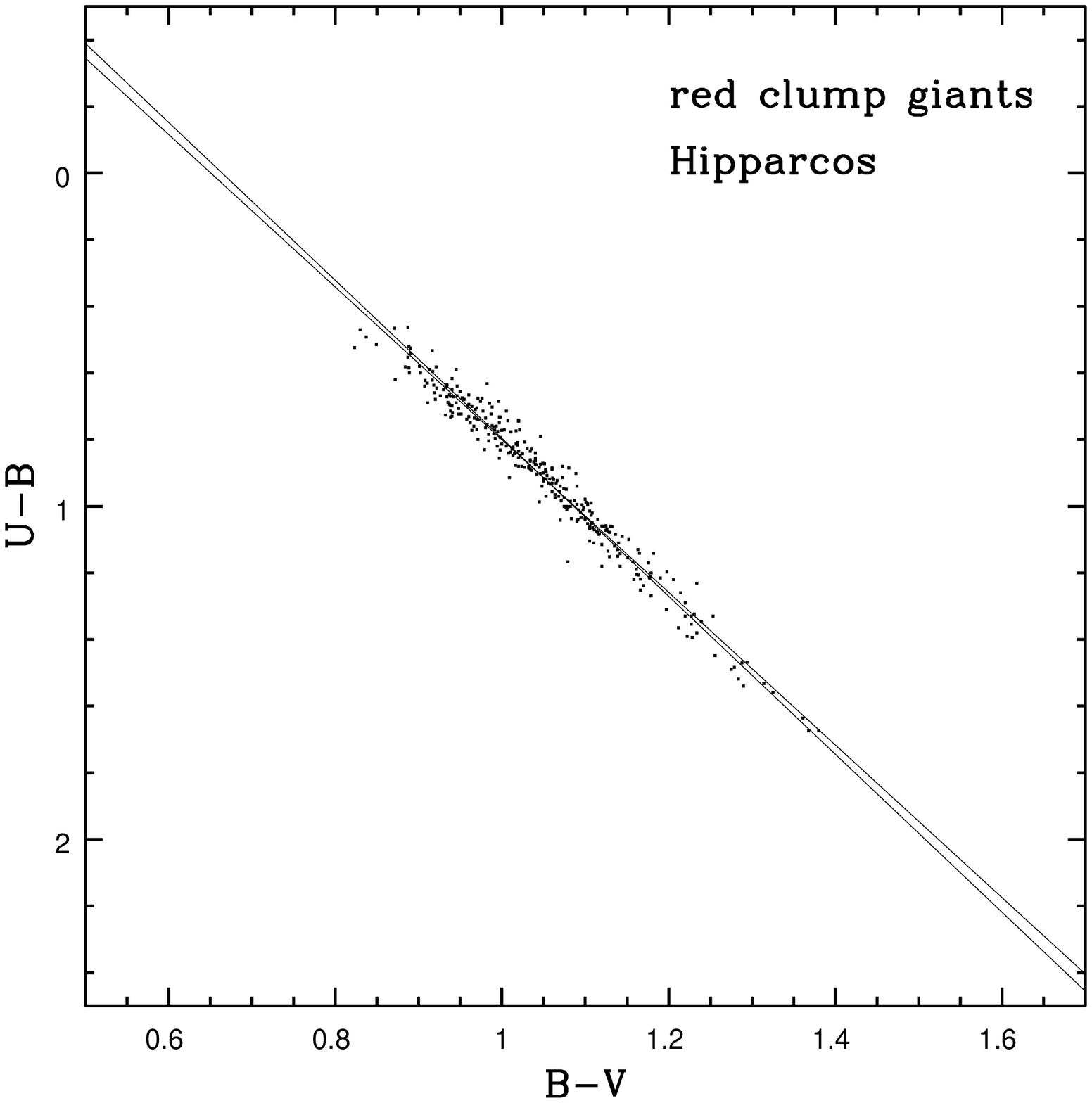,bbllx=20pt,bblly=145pt,bburx=550pt,bbury=680pt,width=8cm,clip=}
\vspace*{3pt}
\FigCap{Color--color diagram: ${(B-V)-(U-B)}$ for 308 red clump giants for 
which Hipparcos measured parallaxes with the accuracy better than 10\%. The 
two thin lines that cross are the regression lines of one color with respect 
to another.} 
\end{figure}

\Section{Discussion}
\Subsection{{\it UBVI} Observations}
A comparison of several hundred red clump giants for which OGLE-I
photometry  (Szyma{\'n}ski \etal 1996) and the current OGLE-II
photometry are available  allowed us to determine a small offset between
the two. The OGLE-I $V$-band is 0.021~mag fainter, the $I$-band is
0.035~mag brighter, and the ${(V-I)}$  color is 0.056~mag redder than
the corresponding quantities determined with  the OGLE-II, which we
consider to be more accurate.

The average colors for the red clump giants presented in Table~2, and
the {\it  rms} deviations of the stars from the average are: (1.114,
0.117), (1.081,  0.191), and (1.148, 0.510) for ${(V-I)_0}$,
${(B-V)_0}$, and ${(U-B)_0}$,  respectively. The corresponding numbers
for the Hipparcos red clump giants  listed in Table~3 are: (1.021,
0.091), (1.050, 0.102), and (0.916, 0.239) for  these colors. Notice
that the range of colors for Baade's Window stars is  larger than for
Hipparcos stars, and the difference increases strongly toward short
wavelengths,  being the largest in ${(U-B)_0}$, as expected for a
population with a larger  range of metallicities.

We may compare our mean ${(V-I)_0}$ colors and their range with the 
corresponding numbers given by Paczy{\'n}ski and Stanek (1998): (1.16,
0.14)  for Baade's Window (based on OGLE-I data but corrected to the
OGLE-II  photometric system) and (1.01, 0.08) for the Hipparcos stars.
Our new  Hipparcos numbers are practically the same as those of
Paczy{\'n}ski and  Stanek (1998), but there is some difference for
Baade's Window stars. There  may be several reasons: Paczy{\'n}ski and
Stanek (1998) used a much larger  area in their study, with a larger
range of interstellar extinction.  OGLE-I photometric errors were larger
which may be responsible for the  apparent larger range of colors, in
addition to the systematic offset between  OGLE-I and OGLE-II, which has
been allowed for.

The data we present in this paper cannot by itself resolve the current 
ambiguity about the relation between the colors and the metallicity of
red  clump giants (Paczy\'nski 1998). However, the catalogs of these
stars in  Baade's Window (Table~2) and near the Sun (Table~3) can be
used to select  stars over the full range of colors for spectroscopic
studies. 

Tight color--color relations apparent for the Hipparcos stars (Figs.~8
and  9) clearly indicate that either there is a single parameter that
determines  the red clump colors, or there is a degeneracy between
several parameters,  which conspire to change stellar colors in the same
direction in the three  dimensional space of ${(V-I)-(B-V)-(U-B)}$. The
same appears to be true for  the red clump giants in Baade's Window
(Figs.~6 and 7). The scatter is larger  in Baade's Window for at least
two reasons: photometric errors are larger than  for the bright
Hipparcos stars, and reddening corrections are not perfect. It  is
somewhat disturbing that the slope of the ${(B-V)_0-(V-I)_0}$ relation
is  somewhat different for the stars in Baade's Window than those near
the Sun. In  principle there is a possibility that the reddening vector
varies  significantly with the stellar color, but we did not make an
estimate of this  effect. 

If future spectroscopic determinations of the metallicity establish only
a  weak correlation with colors of red clump giants, then it will be
clear that  another parameter, most likely the mass (\ie age) of a star
affects its  colors too. This would imply that red clump giants make
large loops in the CMD  during their core helium burning phase. However,
if it turns out that the  spectroscopic metallicity is very well
correlated with colors, then the  implication will be that the
dependence on mass (age) is weak, and the  evolutionary loops are small.

Note, that our samples of the red clump giants are not likely to be
biased by  their colors, as they are all well above the detection limit,
except for the reddest stars, for which {\it U}-band magnitude was close
to the detection limit (\cf Fig.~1).

Following the approach of Paczy\'nski and Stanek (1998) we determined that the 
number density of red clump stars peaks at ${I_0=14.37\pm0.02}$. Of course, 
the true error may be as large as 0.1 mag because the zero point of the 
interstellar extinction is uncertain (Alcock \etal 1998, Gould \etal 1998). 
For comparison, Paczy{\'n}ski and Stanek (1998) obtained ${I_0=14.36}$ for the 
peak of the red clump stars, allowing for the systematic offset between OGLE-I 
and OGLE-II. 

Stanek \etal (1999) have presented {\it UBVI} photometry in a nearby
field BW8  in Baade's Window. Their results are in substantial agreement
with ours,  although there appear to be some systematic differences of
order of 0.05~mag  in zero points of photometry. They also discuss the
effect of metallicity on  colors of the red clump stars. 

\Subsection{Stellar Models}
It is remarkable how well the models of red clump giants agree with the 
observations. Recently, evolutionary tracks for a broad range of
metallicities  and masses were calculated by Bertelli \etal (1994),
Jimenez, Flynn and  Kotoneva (1998), and by Girardi \etal (1998). The
results are best summarized  by Fig.~1 of Girardi \etal (1998), which
clearly shows that the absolute {\it  I}-band magnitude changes very
little with stellar mass (\ie age), but it decreases by 0.2--0.3~mag
when the metallicity decreases from ${Z=0.03}$ to  ${Z=0.001}$, in a
fairly good agreement with the Udalski's (1998b) empirical  calibration.
Models demonstrate that while the metallicity is the dominant  factor
affecting the colors, the stellar mass (age) makes a non-negligible 
contribution, especially at high metallicities, where red clump giants
make  fairly large horizontal loops in the CMD. This suggests that there
is some  degeneracy, in the sense that metallicity and stellar mass
(age) move a star  along nearly the same, approximately horizontal
vector in the CMD, and along  approximately the same vector in a
color--color diagram. 

Things are different at very low metallicities, like those found in
Leo~I  (Caputo \etal 1999), where ${Z=0.0004}$ or even 0.0001. Under
these extreme  conditions the red clump appears like a small arc, with
the top and bottom  relatively blue, and the middle relatively red. The
range of luminosities of  the red clump giants in Leo~I is larger than
the range observed in the  Galaxy (\eg Paczy\'nski and Stanek
1998), Magellanic Clouds (\eg Udalski \etal  1998a, Stanek, Zaritsky and
Harris 1998), or in M31 (Stanek and Garnavich  1998). This is in
excellent agreement with the prediction of stellar  evolutionary models
(Bertelli \etal 1994), and indicates a very large spread  in masses
(ages) of the extremely metal poor population observed in Leo I  (Caputo
\etal 1999). 

Things are also different for very young red clump giants, as it is
apparent  in the data (Udalski \etal 1998a) and in the models (\cf
Fig.~1 of Girardi  \etal 1998): the stars younger than 2~Gyr are
brighter than those in the age  group of 2--10~Gyr. However, it is
remarkable, that in all objects with a  large number of red clump giants
that were studied so far: Hipparcos stars and  the Galactic bulge
(Paczy\'nski and Stanek 1998), LMC, SMC, Carina dwarf  galaxy (Udalski
\etal 1998a, Udalski 1998a), M31 (Stanek and Garnavich 1998),  the red
clump shows very little spread in the {\it I}-band magnitude, 
indicating that the stellar population is dominated by stars which are 
2--10~Gyr old. 

While one may argue about 0.1~mag or even 0.2~mag discrepancy between
the  models and observations (\eg Girardi 1999), the situation is
comparable to, or better  than, the situation with the RR~Lyr stars or
with Cepheids, except those  stars do not have empirical calibration as
good as the red clump giants. A  thorough comparison between the models
and the data for the red clump giants  can be found in Stanek \etal
(1999).

While the evolutionary models are in a reasonable agreement with
observations, one should be aware of several problems with the models
which  cannot be readily resolved and which make a perfect agreement
with the data  almost impossible. 

It is well known that theory cannot predict what should be the value of
the  mixing length parameter which is used to describe non-adiabatic
convection  below stellar photosphere. Moreover, the model effective
temperature and  radius are very sensitive to this parameter, and in
practice the calibration  is empirical. There is no reason for the
mixing length parameter to  be a universal constant, and today's theory
cannot make a firm prediction of  the effective temperature of any cool
star, including red clump giants. 

In order to be compared with observations, the effective temperature
of a  stellar model is converted to a color, like ${(V-I)}$, which
introduces next  theoretical step: model atmosphere. While these are
ever more sophisticated  and presumably accurate, ultimately the
accuracy of the $T_{\rm eff}$ \vs  color transformation has to be
empirically verified. The same is true for the  bolometric corrections.
Note that the bolometric luminosity provided by  stellar models is not
sensitive to the value of mixing length parameter.  However, the
bolometric correction rests on model atmospheres and their  empirical
calibration. 

Stochastic mass loss from red giant stars is believed to be responsible
for  the morphology of the horizontal branch (Rood 1973, Chiosi,
Bertelli, and  Bressan 1992). In other words, the amount of matter lost
affects stellar  radius and effective temperature, affecting the size of
a loop the star makes  in the CMD. Red clump giants have identical
interior structure as classical  horizontal branch stars and therefore
their loops are also affected by the  amount of mass loss. As the reason
for the stochastic nature of mass loss is  not known, the size of those
loops cannot be firmly predicted by the theory. 

There is yet another major uncertainty of the models, which is not very
well  known, or not very well remembered. Several decades ago it was
established  that low mass stars develop large semiconvective zones just
outside of their  convective helium burning cores (Schwarzschild 1970,
Iben 1974, and references  therein). The question: "what is the proper
criterion for a marginally stable  gradient of chemical composition" was
debated. There has been no discussion of  this subject lately, and there
is no reference to the semiconvection in recent  papers on models of
horizontal branch or red clump stars. It is not clear  whether the
problem has been solved or forgotten. 

Another very delicate issue is the extent to which chemical composition 
profile affects the size of the loops that stars make in the CMD while
burning  helium in their cores. Robertson (1971), and Lauterborn,
Refsdal, and Weigert  (1971) demonstrated that even very small changes
in the stellar interior had a  huge effect on the loop size. This is not
important when the red clump models  are very close to the red giant
branch, \ie when the loop is very small, and  the range of colors is
very small. However, high metallicity clump models  appear to have
fairly extended loops (\cf Fig.~1 of Girardi \etal 1998), and  the size
of these may be very sensitive to many details of stellar structure. 

The issue of the loop size can be most reliably resolved with the
spectral  observations, which may establish how tight (or not tight) is
the correlation  between the colors and metallicity of the red clump
giants. The tighter the  correlation, the smaller the loops have to be,
and {\it vice versa}.

Considering the many difficulties with the stellar models it is not
likely  that they can provide quantities like $M_I$ and ${(V-I)_0}$ with
a precision  better than 0.1--0.2~mag, unless empirically calibrated.
This is roughly the  level at which the models and the observations
appear to agree or disagree  with each other. 
\begin{figure}[htb]
\hglue2cm\psfig{figure=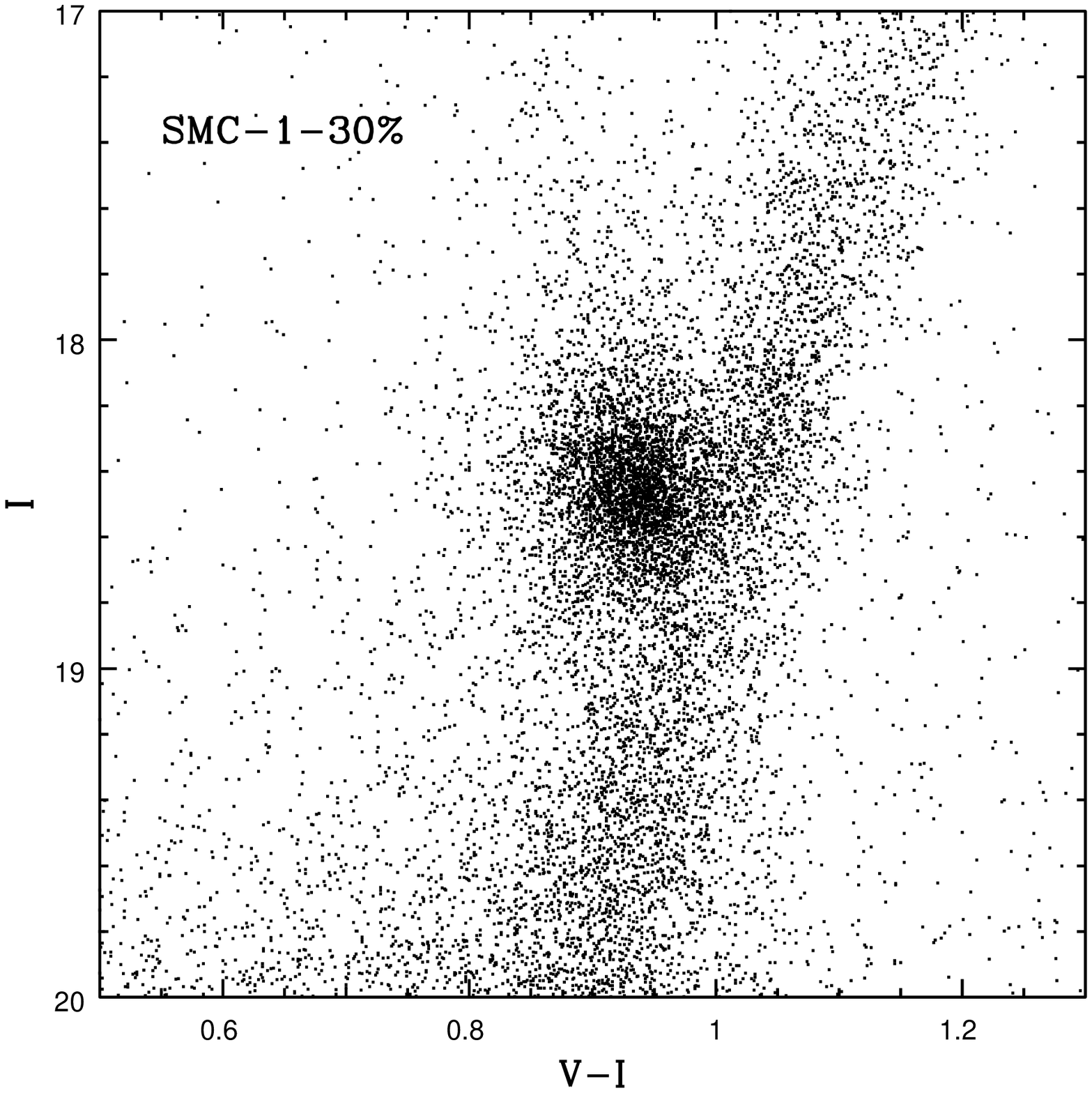,bbllx=20pt,bblly=150pt,bburx=550pt,bbury=680pt,width=9cm,clip=}
\vspace*{3pt}
\FigCap{Color--magnitude diagram for 30\% of stars in the SMC field SMC$\_$SC1 
(Udalski \etal 1998b). Note how narrow is the red giant branch and how compact 
is the red clump in this field with low and uniform reddening.} 
\end{figure}

\Subsection{Red Clump in the SMC}
Although the Small Magellanic Cloud is not the topic of this paper, we
present  in Fig.~10 a color magnitude diagram for the red clump for a
region near the  south-west end of the SMC bar. We took the data from
the SMC Photometric Maps  published by the OGLE team (Udalski \etal
1998b), which have been accessible  from the OGLE archive for over a
year. The main reason for presenting Fig.~10  is to show how compact the
red clump is when the reddening is not a problem,  the range of
distances is small, and the range of metallicities is low too.  The data
from the SMC maps may be useful to test and/or calibrate theoretical 
models. In particular, it is striking that the red clump in the SMC has
no  apparent structure, except for a cloud of stars above the main body
of the  clump; these are likely the younger and more massive stars as
shown in Fig.~1  of Girardi \etal (1998). 

\Acknow{It is a great pleasure to acknowledge numerous discussions with
Dr.\  K.Z.~Stanek. We are very indebted to Dr.\ J.-C.\ Mermilliod for
providing us  with the {\it UBV} data for the Hipparcos stars in a
computer readable form.  We are  very grateful to Dr.\ P.\ Popowski for
pointing out some important omissions  in the original version of this
paper. We are also grateful to Dr.\ J.\ Cohen for pointing out mistakes
in the original Table~3. This work was supported with the NSF grants
AST--9530478 and AST--9820314 to  B.\ Paczy{\'n}ski and the Polish KBN
grant 2P03D00814 to A.\ Udalski.}

\end{document}